\documentstyle[aps,prl,preprint,epsfig]{revtex}
\begin{document}

\title{Two-neutron separation energies, binding energies and phase
transitions in the interacting boson model}
\author{J.E.~Garc\'{\i}a-Ramos$^{1,2,}$\footnote{Visiting 
postdoctoral fellow of the Fund of Scientific Research, Flanders, 
Belgium.}, C.~De Coster$^{1,}$\footnote{Postdoctoral fellow 
of the Fund of Scientific Research, Flanders, Belgium.}, 
R.~Fossion$^{1}$ and, K.~Heyde$^{1}$}

\address{$^1$Institute for Theoretical Physics, Vakgroep Subatomaire
en Stralingsfysica, Proeftuinstraat 86, B-9000 Gent, Belgium}
\address{$^2$Departamento de F\'{\i}sica At\'omica, Molecular y Nuclear,
  Universidad de Sevilla, Apartado 1065, 41080 Sevilla, Spain}
\date{\today}
\maketitle

\begin{abstract}
In the framework of the interacting boson model the three transitional
regions (rotational-vibrational, rotational--$\gamma$-unstable and, 
vibrational--$\gamma$-unstable transitions) are reanalyzed. A new kind 
of plot is presented for studying phase transitions in finite systems 
such as atomic nuclei. The importance of analyzing binding energies
and not only energy spectra and electromagnetic transitions, describing 
transitional regions is emphasized. We finally
discuss a number of realistic examples. 
\end{abstract}

\vspace{2cm}
\newpage


\section{Introduction}
\label{intro}
In the last few years, interest for the study of phase transitions and
phase coexistence in atomic nuclei has been revived 
\cite{Rowe98,Iach98,Cast99,Joli99} in particular making use of the
Interacting Boson Model (IBM) \cite{Iach87}. 
In the chart of nuclei, three transitional regions can be
distinguished where
one observes rapid structural changes:  
(a) In the Nd-Sm-Gd region, one observes a 
change from spherical to well-deformed nuclei when 
moving from the lighter to the heavier isotopes; in the IBM
language this is the $U(5)$-$SU(3)$ transitional region; (b) in the
Ru-Pd region, one notices that the lighter isotopes are spherical
while the heavier ones indicate a  $\gamma$-unstable character, 
this is the $U(5)$-$O(6)$ region; (c) in the Os-Pt region, 
the lighter isotopes are well deformed while the heavier shown
$\gamma$-unstable properties, this is the $SU(3)$-$O(6)$ transitional
region. Although these three transitional regions have been studied 
extensively in the framework of the IBM 
\cite{Scho78,Scho80,Stac82,Cast78}
(note that in this paper we use the simplest version of the
model, IBM-1, therefore we do not consider the
transitional regions using IBM-2 \cite{Scho80,Isac80,Bijk80}), 
the discussion of phase 
transitions has not always been treated in a proper way. In
particular one of the weak points is how to define an appropriate
control parameter in a system such as the atomic nucleus where this
parameter is fixed (for given $N$ and $Z$) and cannot be controlled 
externally \cite{Diep80}. 
This problem has been recently considered in a study by Casten {\it et
al.~} \cite{Cast99}.     

Binding energy (BE) serves as an appropriate ``signature'' for phase
transitions in nuclei \cite{Diep80,Feng81}.
The binding energy of the nucleus or, equivalently its mass, is
one of the more important nuclear properties to be determined. 
In some cases this is the only knowledge about a nucleus when 
situated far from stability. In the last few years,
the development of new experimental facilities has given access to very
unstable nuclei and many new mass measurements have been  
obtained \cite{Schw00}. In the light of
the variety of nuclear properties (nuclear mass, nuclear moments,
$\ldots$) that have been obtained recently, a theoretical study
concentrating on the nuclear binding energy and, related to that,
possible phase transition is desirable. 

Binding energies contain important information about nuclear
structure. In IBM calculations, only
spectra and electromagnetic transitions used to be analyzed and the
description of the binding energy is in most 
of the cases excluded. In
the present article we show that it is  convenient to explore nuclear
binding energies in order to obtain a consistent description, especially
when one is dealing with chains of isotopes. It must be noted that  
the study of
binding energies can help to discriminate between apparently
equivalent descriptions of energy spectra and electromagnetic 
transitions rates for a given nucleus or, even for a chain of nuclei. 

In section \ref{ph-trans} we review the study of phase transitions 
in the IBM. In section \ref{id-ph-trans} a new way of dealing 
with phase transitions in nuclei will be presented together with 
the study of two-neutron separation energies (S$_{2n}$) in regions 
where phase transition might happen. In section \ref{real} we study some
realistic cases and, finally, in section \ref{conclu}, we present our 
conclusions.

\section{Transitional regions and phase transitions in the IBM: a
study of nuclear binding energies}
\label{ph-trans}
The IBM is an algebraic model able to describe the low-lying
collective states in even-even nuclei in terms of bosons which
carry either angular momentum $L=0$ ($s$ bosons) 
or angular momentum $L=2$ ($d$ bosons) \cite{Iach87}. 
The system of bosons, for which the number
of bosons equals
half the number of valence fermions, $N=n/2$, interacts through a
Hamiltonian that typically includes up to two-body interactions,
is number conserving and rotationally invariant. The original version
of the model, the IBM-1, is used in the present article.

The IBM has been extensively used during the last decades for the
study of medium and heavy nuclei \cite{Cast88}. 
One of the main challenges for the model
was the description of transitional regions, where the structure of
the nuclei can change rapidly along a chain of isotopes. In particular,
different transitional regions have been obtained where such changes
in the nuclear structure appear: from spherical to well deformed
nuclei (Nd, Sm, and Gd region), from
well deformed to $\gamma$-unstable nuclei (Ru and Pd region) or 
from spherical to $\gamma$-unstable (Os and Pt region). In the IBM 
language a spherical nucleus is related to the $U(5)$ limit, 
a well deformed nucleus is
related to the $SU(3)$ limit, while $\gamma$-unstable nuclei
correspond to the $O(6)$ limit. A  compact Hamiltonian for 
analyzing the three previous regions can be depicted as follows,
\begin{equation} 
\label{ham1}
\hat H=\kappa \Big(N {1-\xi\over\xi}\hat n_d-
        \hat Q\cdot\hat Q\Big)+\kappa' \hat L\cdot\hat L, 
\end{equation}
with $0\leq\xi\leq1$.  
The coefficient of the one-body term can be 
rewritten as $N(1-\xi)/\xi=\epsilon_d/\kappa$, where $\epsilon_d$ is
the single-particle energy for the $d$ bosons. On the other hand, 
$\hat {n}_d$ is the $d$ boson number operator and
\begin{eqnarray}
\label{L}
\hat L&=&\sqrt{10}(d^\dag\times\tilde{d})^{(1)},\\
\label{Q}
\hat Q&=& s^{\dagger}\tilde d
+d^\dagger\tilde s+\chi(d^\dagger\times\tilde d)^{(2)}.
\end{eqnarray} 
The symbol $\cdot$ represents the scalar product. In this work the
scalar product of two operators with angular momentum $L$ is defined as 
$\hat T_L\cdot \hat T_L=\sum_M (-1)^M \hat T_{LM}\hat T_{L-M}$ where 
$\hat T_{LM}$ corresponds to the $M$ component of the operator 
$\hat T_{L}$. The operator 
$\tilde\gamma_{\ell m}=(-1)^{m}\gamma_{\ell -m}$ (where $\gamma$
refers to $s$ and $d$ bosons) is introduced to ensure the
tensorial character under spatial rotations.
With this Hamiltonian one can explore the
three transitional regions which correspond to the segments of the
symmetry triangle \cite{Cast81}.

One of the most striking facts that happen in the transitional regions 
is the possibility of observation of phase transitions. A phase
transition can roughly be defined as a qualitative change in a given
property of a system. In order to characterize the properties of the
system one has to introduce an order parameter. 
Another important concept is the order of a transition which is the degree
of the derivative for a given experimental observable which first
experiences a discontinuity. In the atomic nucleus this experimental
observable is the binding energy or, the related two-neutron separation
energy (S$_{2n}$). Although phase transitions are
strictly well defined in macroscopic systems and the atomic nucleus is 
a finite system, a number of studies have shown that the concept of a
phase transition retains its validity and usefulness in small systems
too \cite{Gilm78,Gilm78b,Feng81,Gilm81}. 
A very useful tool in order to discuss phase transitions in finite system
are the coherent states which, in the case of the IBM, are also
known as intrinsic states \cite{Gino80,Diep80,Garc98c}. 
The intrinsic-state formalism provides a
connection between the IBM and the Geometrical Collective Model 
\cite{Bohr77}. The
key point for establishing this connection is to consider that the
dynamical behavior of the system can be described, to a first
approximation, as arising from independent bosons moving in an average
field \cite{Duke84}. 
The ground state of the system is a condensate, $|c\rangle$, of
bosons that occupy the lowest-energy phonon state
$\Gamma^\dag_c$,
\begin{equation}
\label{GS}
| c \rangle = {1 \over \sqrt{N!}} (\Gamma^\dagger_c)^N | 0 \rangle,
\end{equation}
where
\begin{equation}
\label{bc}
\Gamma^\dagger_c = {1 \over \sqrt{1+\beta^2}} \left (s^\dagger + \beta
\cos     \gamma          \,d^\dagger_0          +{1\over\sqrt{2}}\beta
\sin\gamma\,(d^\dagger_2+d^\dagger_{-2}) \right).
\end{equation}
Here, the variables $\beta$ and $\gamma$, that are obtained after
minimization of the ground-state energy, are the order parameters
of the system. A spherical nucleus will have as order parameters
$\beta=0$ and an arbitrary value for $\gamma$; a deformed nucleus is
characterized by a finite value of $\beta$ (in the $SU(3)$ limit for
$N\rightarrow\infty$ one has $\beta=\sqrt{2}$) 
and $\gamma=0$ (prolate nucleus) or $\gamma=\pi/3$
(oblate nucleus); and a $\gamma$-unstable nucleus corresponds to
$\beta=1$ (in the limit $N\rightarrow\infty$) 
and to an arbitrary value for $\gamma$. A phase transition
will produce changes in the order parameter of the system. Using the
intrinsic state we get ``microscopic'' information on the order
parameter.  

\begin{figure}[]
\begin{center}
\mbox{\epsfig{file=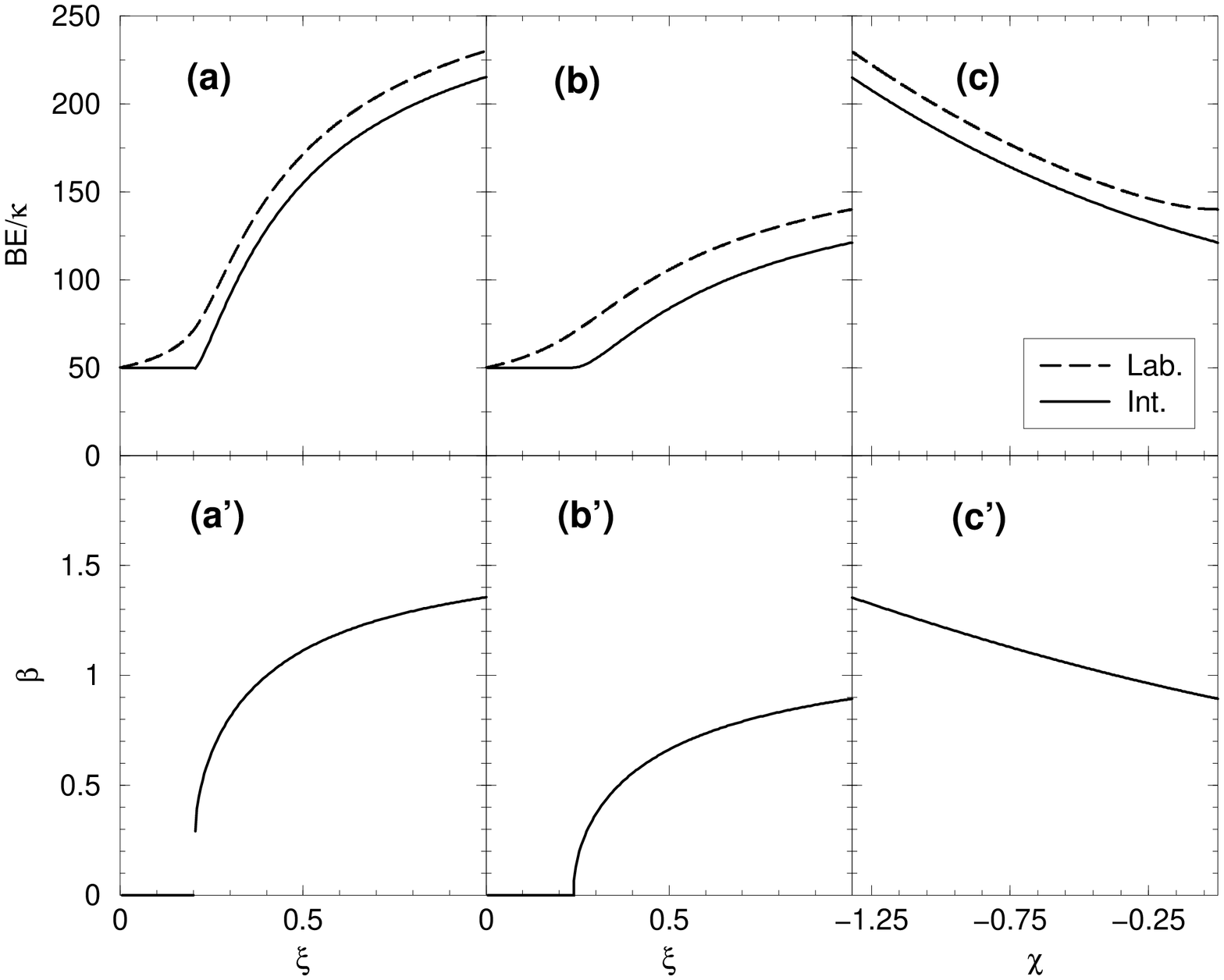,height=14.0cm,angle=-90}}
\end{center}
\caption{Binding energies and $\beta$ as a function of the
control parameter, in the transitional
regions (a) $U(5)$-$SU(3)$, (b) $U(5)$-$O(6)$, and (c) $SU(3)$-$O(6)$.
Dashed lines correspond to laboratory calculations while full
lines correspond to intrinsic state results. Number of bosons $N=10$.}
\label{fig-trans-be}
\end{figure}

First, we analyze the behavior of the binding energy and two-neutron
separation energy in the three transitional regions. In the three
cases the results are independent of the value of $\kappa'$ in
eq.~(\ref{ham1}). For the $U(5)$-$SU(3)$ transition the value of
$\xi$ varies from $0$ ($U(5)$) to $1$ ($SU(3)$) with
$\chi=\pm\sqrt{7}/2$ \cite{Cast81};
in the $U(5)$-$O(6)$ transition the value of $\xi$ 
varies from $0$ ($U(5)$) to $1$ ($O(6)$) with $\chi=0$ \cite{Cast81};
finally in the transition $SU(3)$-$O(6)$ the value of $\chi$ varies from
$\pm\sqrt{7}/2$ ($SU(3)$) to $0$ ($O(6)$) with $\xi=1$ \cite{Cast81}. 
In figures \ref{fig-trans-be}a, \ref{fig-trans-be}b, 
and \ref{fig-trans-be}c we
present the binding energy in the laboratory system and in the 
intrinsic frame ($BE/\kappa$) while in figures \ref{fig-trans-be}a',
\ref{fig-trans-be}b', and \ref{fig-trans-be}c' we present 
the value of $\beta$ in the three  transitional regions for a number 
of bosons $N=10$. 
First of all, it must be emphasized that the
binding energy calculated in the intrinsic system is always below the
exact value because it derives from a variational method. One can also 
realize that the intrinsic-state calculations show a sharp behavior
in the phase transition regions while the diagonalization in the
laboratory frame leads
to smooth curves. In this sense the variational procedure mimics very
well what happens in the thermodynamic limit. One of the
characteristics of phase transitions in macroscopic systems is that
they are very sharp, but in finite system the sharpness will depend on
the number of particles.  

In figure
\ref{fig-trans-be}a, a first order transition shows up (discontinuity in
the slope of the BE), and also a discontinuity in the value of the
order parameter exits. In figure \ref{fig-trans-be}b, a phase
transition also appears, but in this case it is a second-order transition
(discontinuity in the curvature of the BE).
One observes, more generally, that there exits a line of first order 
transitions for Hamiltonians with 
$\chi\neq 0$ and just one point for a second-order transition  
in the case of $\chi=0$ \cite{Iach87}. 
The parametrization of the Hamiltonian
(\ref{ham1}) is very appropriate because the phase transition  
appears approximately at the same value of the control parameter ($\xi$) for
any value of $N$ \cite{Joli99}. 
This critical point, $\xi_c$, can be  obtained easily
from the expression for the energy given in \cite{Isac81} as
\begin{equation} 
\label{xic}
\xi_c={N\over 5N +\chi^2-8}.
\end{equation}  
For $\xi<\xi_c$ the potential has a local minimum in $\beta=0$, while
for $\xi>\xi_c$ this local minimum becomes a local maximum.
It is clear from this expression that for a large number of bosons,
$\xi_c\approx0.2$.
Finally, in figure \ref{fig-trans-be}c no phase
transition appears and both calculations, exact diagonalization and
intrinsic-state, lead to a smooth transitions.

\begin{figure}[]
\begin{center}
\mbox{\epsfig{file=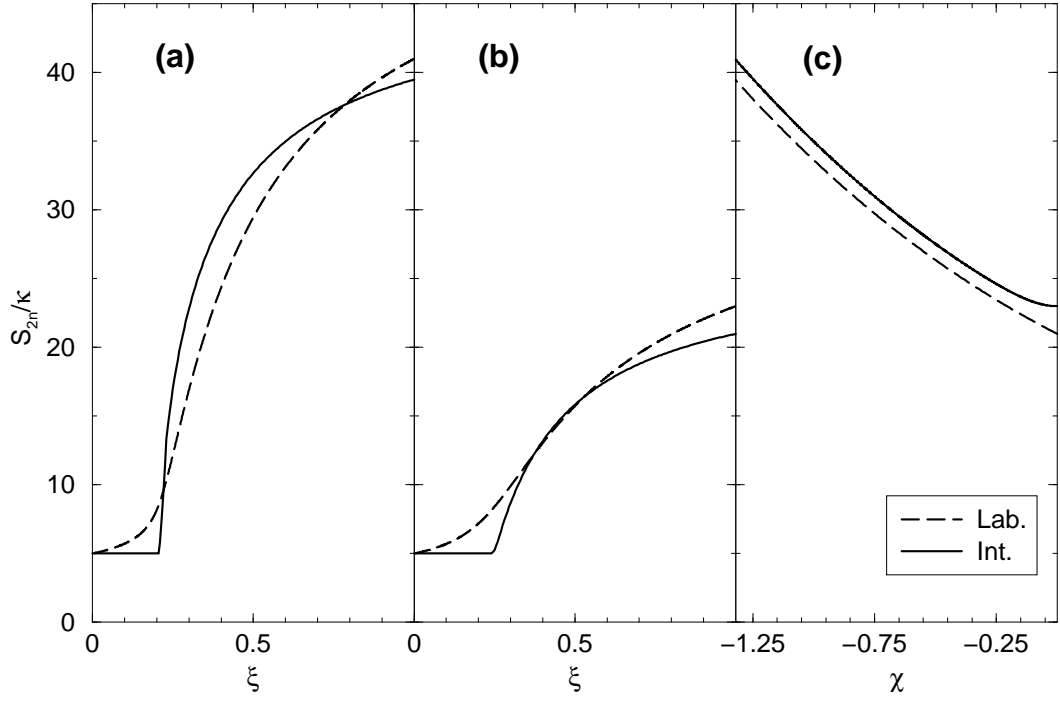,height=14.0cm,angle=-90}}
\end{center}
\caption{Two-neutron separation energies (S$_{2n}$) as a function of the
control parameter, in the transitional
regions (a) $U(5)$-$SU(3)$, (b) $U(5)$-$O(6)$, and (c) $SU(3)$-$O(6)$. 
Dashed lines correspond to laboratory calculations while full
lines correspond to intrinsic state results. Number of bosons $N=10$.}
\label{fig-trans-s2n}
\end{figure}

In a next step, we analyze the behavior of the two-neutron separation
energy. The definition that will be used along this paper is,
\begin{equation}
\label{s2n}
S_{2n}(N)=BE(N)-BE(N-1). 
\end{equation}
In the calculation of S$_{2n}$ two different nuclei are involved, which
makes necessary to impose an assumption 
in order to relate the Hamiltonians for
the nuclei with $N$ and $N-1$ bosons. Here, we consider the
simplest (not unrealistic) case which implies the same Hamiltonian 
in both nuclei. In figure \ref{fig-trans-s2n} we study the same
transitions as in figure \ref{fig-trans-be} and essentially the same 
remarks hold. Note in this case the crossing of
intrinsic-state and laboratory calculations, which is allowed.

So, in this section we have reviewed the study of phase transitions in
atomic nuclei, in particular using the IBM, and have shown it is 
straightforward and clear from a theoretical point of view. 
In this analysis, however, there is a
weak point: the control parameter $\xi$ ($\chi$) is not a genuine one,
because it cannot be modified externally. The control parameter only
changes when one is moving along a chain of nuclei and also assumes a
change in the number of bosons. These two problems have already been
remarked and alternative solutions have been proposed \cite{Cast99}. 
In the next sections, we propose a new strategy for attacking 
this problem.   

\section{How to identify phase transitions}
\label{id-ph-trans}

\subsection{The new diagrams ``binding energy-number of bosons-control 
parameter''} 
\label{diagr}
A consistent treatment of phase transitions in nuclei must go beyond
the study carried out in section \ref{ph-trans} because one has to treat
simultaneously the control parameter $\xi$ ($\chi$) 
and the number of bosons
$N$. It must be emphasized that $N$ is not a control parameter because
it is a discrete variable and it is fixed in each individual nucleus. 
Strictly speaking $\xi$ ($\chi$) is neither a genuine control
parameter, because its value is also fixed in each given nucleus, nor
can it be controlled externally. Only the appropriate interplay of $N$ 
with $\xi$ ($\chi$) can lead to a proper definition of an order 
parameter, although with discrete values.

The appropriate way for treating phase transitions and
transitional regions is to plot in the same figure the value of the
binding energy versus the control parameter and this for 
different values of $N$. In such a plot, a transition will develop 
through changes in the control parameter, but at the same time going 
through curves with a different number of bosons. 
In order to illustrate this new procedure, we discuss  the transitional
regions $U(5)$-$SU(3)$, $U(5)$-$O(6)$ and, $SU(3)$-$O(6)$ in 
figures \ref{fig-be-n}a, \ref{fig-be-n}b, and \ref{fig-be-n}c
respectively. In these figures, only the laboratory results (IBM
diagonalization) are presented. 
As an example, trajectories for real nuclei are also plotted in 
each figure. A procedure in order to obtain the value of the control 
parameter, for each nucleus, will be explained in
section \ref{real}. The more interesting cases
appear in figures \ref{fig-be-n}a and \ref{fig-be-n}b because there,
phase transitions indeed happen. So, looking at these figures, one can
easily see if a given nucleus is situated in the spherical region 
($\xi$ near to $0$), in the deformed region ($\xi$ around 1) 
or at the critical point.   

One might be tempted to consider similar plots for the two-neutron
separation energies. In this case one must be very careful because
the Hamiltonian for $N$ and $N-1$ bosons are different. That means
that S$_{2n}$ is not a function of a single control parameter, $\xi$
($\chi$), anymore, but becomes a function of two control 
parameters, one for $N$ and another one for $N-1$. 

The essential difference between phase transitions in
nuclei compared to macroscopic systems is the existence of a control
parameter in the macroscopic case, such as temperature or pressure,
that can be controlled externally. In the case of atomic nuclei, 
the only way for changing the value of the control parameter 
is to move along a chain of nuclei. It is clear that the variation 
of the control parameter is thereby fixed and cannot be modified 
externally. 
 
\subsection{Crossing the phase transition region: anomalies in the 
two-neutron separation energies}
\label{cross}

In previous sections we have studied the behavior of the binding
energy as a function of the control parameter, {\it i.e.~}as a function
of the Hamiltonian, in the three transitional regions of the IBM. Due
to the specific characteristics of phase transitions in atomic
nuclei, it is not {\it a priori} clear how different observables will 
behave when crossing a critical value through a chain of
isotopes. More particular, one can observe the two-neutron separation
energies as a function of the nucleus in the chain that one 
considers. S$_{2n}$ is indeed a very appropriate observable to be analyzed
because it contains important information about the nuclear structure,
in particular on the ground state.

Before starting the present analysis we emphasize the fact that there 
exits an extra contribution to the binding energy not yet included
when using the Hamiltonian of equation (\ref{ham1}), {\it i.e.~}
\begin{equation} 
\label{be-lin}
E'=E_0+A N + {B\over 2} N (N-1)
\end{equation} 
Those extra terms, deriving from the linear and quadratic $U(6)$
Casimir invariant, take into account the bulk contribution of the nuclear
interaction. Including those extra terms in S$_{2n}$, one obtains a linear
contribution to be added to the IBM calculation, resulting 
\begin{equation}
\label{s2n-lin} 
S_{2n}(N)=(A-B/2)+B N +BE_{IBM}(N)-BE_{IBM}(N-1).
\end{equation} 
When the linear dependence is excluded from S$_{2n}$, we will refer to
that result as S'$_{2n}$. It must be noted that the 
specific nuclear structure IBM contribution in the three
dynamical limits ($U(5)$, $SU(3)$, and $O(6)$) is also linear in the
boson number \cite{Iach87,unpu} and in this context 
only a particular change in the internal structure of
the nuclei, as happens during the phase transitions, can perturb this
linear dependence.

In order to be sure one encounters a phase 
transition or passes through 
a specific transitional region, one has to establish the value of the
control parameter for each nucleus. In the following we will
concentrate on chains of isotopes (fixed number of protons), being the
nuclei characterized by a given number of bosons $N$ or by the 
number of neutron pairs $N_\nu$. As a consequence one has 
to determine the functional relation $\xi=\xi(N)$ ($\chi=\chi(N)$) 
or, in terms of the number of neutron pairs, the relation 
$\xi=\xi(N_\nu)$ ($\chi=\chi(N_\nu)$). Although the
latter functional relation must be determined from experimental data, 
one can use a simple function and see if it is possible to obtain 
some physical insight on the structure of S$_{2n}$. For the present  
study we fix a chain of isotopes with $5$ protons pairs, 
$N_\pi=5$, and a variable number of
neutrons pairs, ranging from $N_\nu=0$ to $N_\nu=10$. Two functional
dependences will be used, firstly linear and secondly quadratic. Next,
we show the different parametrizations for the different transition
regions,
\begin{itemize}
\item $U(5)$-$SU(3)$ and $U(5)$-$O(6)$ transitional regions.
\begin{eqnarray}
\label{xi1}
\xi&=&0.099 N_\nu +0.01,\\
\label{xi3}
\xi&=&0.0099 N_\nu^2 +0.01,
\end{eqnarray}
\item $SU(3)$-$O(6)$ transitional region.
\begin{eqnarray}
\label{chi1}
\chi&=&-{\sqrt{7}\over 20}N_\nu\\ 
\label{chi2}
\chi&=&-{\sqrt{7}\over 200}N_\nu^2.
\end{eqnarray}
\end{itemize}
Using parametrizations (\ref{xi1},\ref{xi3}), $\xi$ ranges from $0.01$ to
$1$. On the other hand, using parametrization (\ref{chi1},\ref{chi2}), 
$\chi$ ranges from $0$ to $-\sqrt{7}/2$. 
In order to establish a realistic parametrization, we start from the
empirical observation that in the $U(5)$-$SU(3)$ transition, the system
passes from the $U(5)$ to the $SU(3)$ limit when the number of bosons is
increasing \cite{Cast88}. Similarly, the transition $U(5)$-$O(6)$
implies that the system goes  from $U(5)$ to $O(6)$ when the number of 
bosons is increasing. Finally, the transition $SU(3)$-$O(6)$ implies 
the system passes from $SU(3)$ to $O(6)$ when the number of bosons is 
decreasing. 

\begin{figure}[]
\begin{center}
\mbox{\epsfig{file=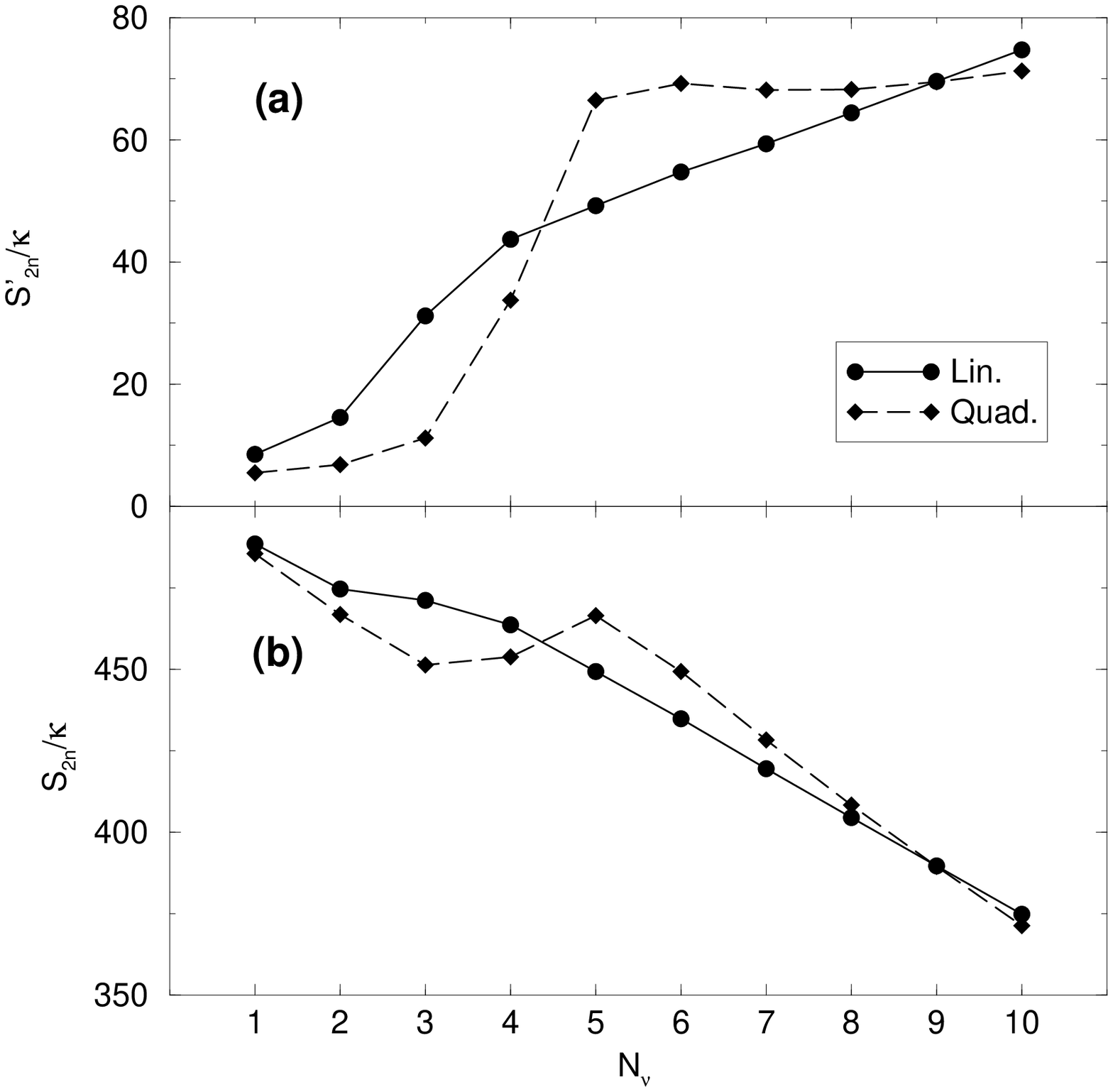,height=14.0cm,angle=-90}}
\end{center}
\caption{Schematic plot of (a) S$_{2n}$ and (b) S'$_{2n}$ in the
$U(5)$-$SU(3)$ transitional region. 
Full lines corresponds to a linear variation of $\xi$
with respect to $N_\nu$ while dashed lines corresponds to a quadratic
dependence.}
\label{fig-s2n-u5-su3}
\end{figure}

\begin{figure}[]
\begin{center}
\mbox{\epsfig{file=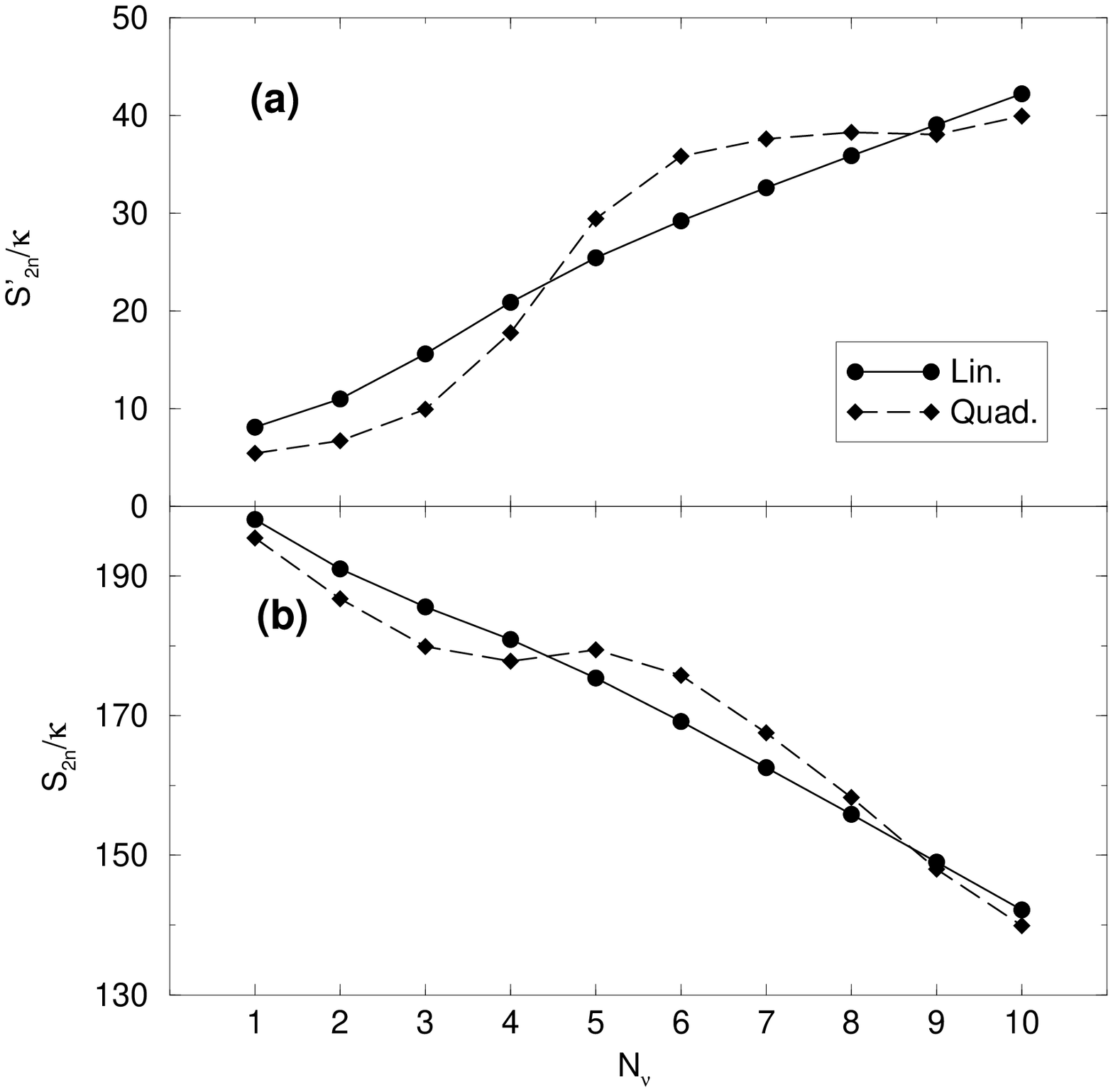,height=14.0cm,angle=-90}}
\end{center}
\caption{Schematic plot of (a) S$_{2n}$ and (b) S'$_{2n}$ in the
$U(5)$-$O(6)$ transitional region. 
Full lines corresponds to a linear variation of $\xi$
with respect to $N_\nu$ while dashed lines corresponds to a quadratic
dependence.}
\label{fig-s2n-u5-o6}
\end{figure}

\begin{figure}[]
\begin{center}
\mbox{\epsfig{file=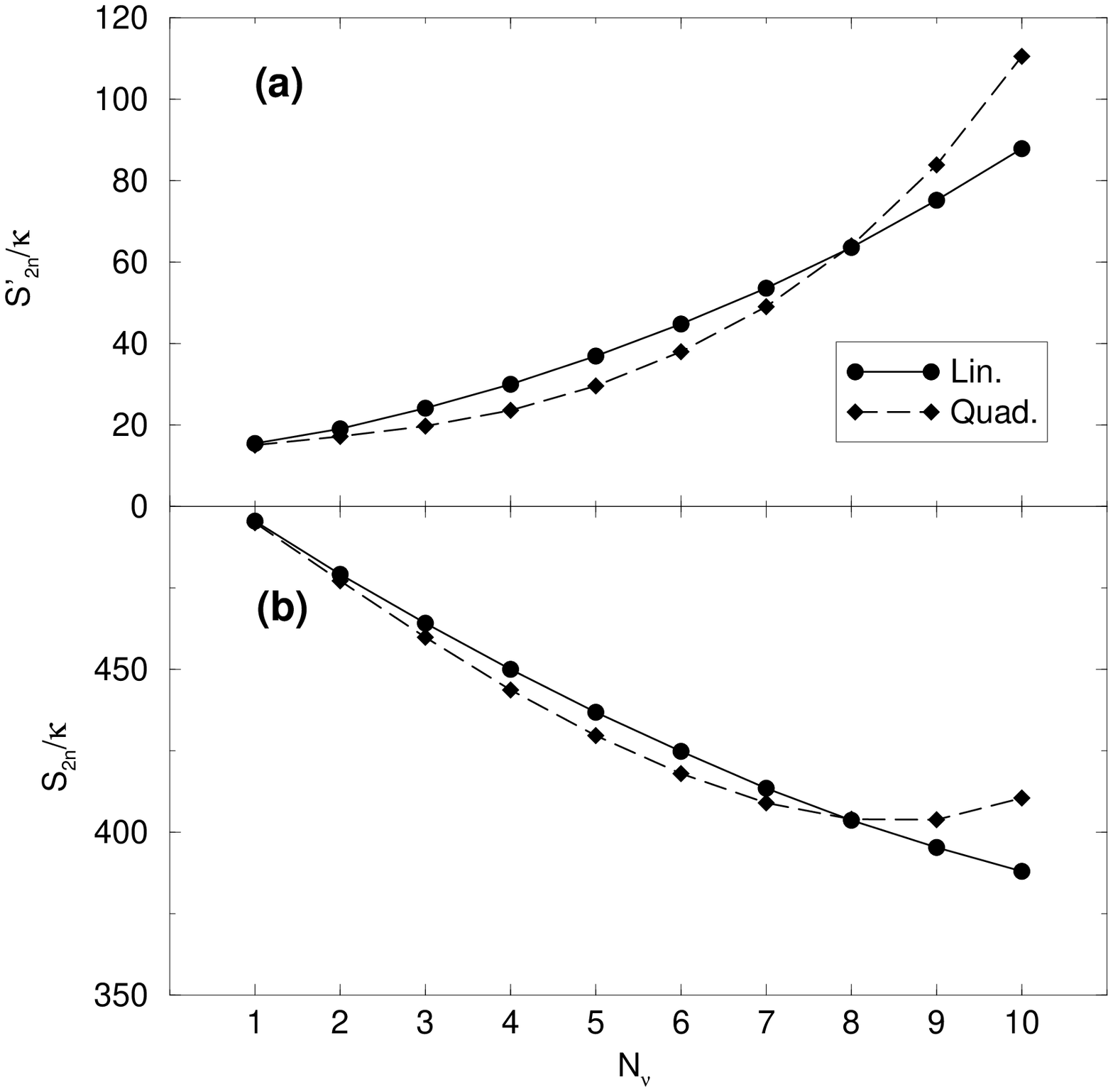,height=14.0cm,angle=-90}}
\end{center}
\caption{Schematic plot of (a) S$_{2n}$ and (b) S'$_{2n}$ in the
$SU(3)$-$O(6)$ transitional region. 
Full lines corresponds to a linear variation of $\chi$
with respect to $N_\nu$ while dashed lines corresponds to a quadratic
dependence.}
\label{fig-s2n-su3-o6}
\end{figure}

In figures \ref{fig-s2n-u5-su3}a and \ref{fig-s2n-u5-su3}b, S$_{2n}$
and S'$_{2n}$ are shown, respectively, for the
$U(5)$-$SU(3)$ transitional region using for $\xi$ the
parametrizations (\ref{xi1},\ref{xi3}). The Hamiltonian (\ref{ham1})
has been used with $\chi=-\sqrt{7}/2$ \cite{Cast81}. 
For convenience, a linear
dependence equal to $S_{2n}^{lin}/\kappa=200-20 N_\nu$ 
has been chosen as a reference value in order to
obtain a realistic behavior in S$_{2n}$. In these figures,
an anomaly in the linear dependence of S$_{2n}$ and
S'$_{2n}$, right at the place where the phase transition happens,  
can be appreciated.
The phase transition point corresponds
approximately to $\xi\approx 0.2$ (see figure \ref{fig-trans-be}a and
\ref{fig-trans-be}a') and taking into account equations 
(\ref{xi1}) and (\ref{xi3})
results a value for $N_\nu$, $N_\nu\approx 2$ and $N_\nu\approx 4$ 
for the linear and quadratic dependence, respectively. 
We point out that in case a
quadratic variation in $N_\nu$ for the control parameter is selected,
the anomaly in S$_{2n}$ is more pronounced than for a linear variation.
In figures \ref{fig-s2n-u5-o6}a and \ref{fig-s2n-u5-o6}b, S$_{2n}$
and S'$_{2n}$ are shown, respectively, for the $U(5)$-$O(6)$ 
transitional region
using for $\xi$ the same parametrization as in the previous case. 
The Hamiltonian (\ref{ham1}) has been used with $\chi=0$
\cite{Cast81}. In this
case a linear dependence equal to $S_{2n}^{lin}/\kappa=200-10 N_\nu$ 
has been used to determine the S$_{2n}$ values. Again, a non-linear 
behavior appears at the point of the phase transition. 
The phase transition point corresponds
approximately to $\xi\approx 0.2$ (see figure \ref{fig-trans-be}b and
\ref{fig-trans-be}b') and taking into account equations 
(\ref{chi1}) and (\ref{chi2})
results a value for $N_\nu$, $N_\nu\approx 2$ and $N_\nu\approx 4$ 
for the linear and quadratic dependence, respectively. 
In this case, a linear variation
in the control parameter leads to an almost linear dependence in
S$_{2n}$ and S'$_{2n}$. Only a control parameter which depends
quadratically on $N_\nu$ produces a kink in the two-neutron separation
energy. Finally, in figures \ref{fig-s2n-su3-o6}a and
\ref{fig-s2n-su3-o6}b, S$_{2n}$ and S'$_{2n}$ are
represented, respectively, for the $SU(3)$-$O(6)$ transitional region 
using for $\chi$ the
parametrizations (\ref{chi1},\ref{chi2}). The Hamiltonian (\ref{ham1})
has been used with $\xi=1$ \cite{Cast81}. 
In this case the following linear
dependence $S_{2n}^{lin}/\kappa=200-20 N_\nu$ has been chosen as a
reference when 
calculating S$_{2n}$. In this transitional region no phase
transition is observed and a smooth curvature result except when one
is approaching the $SU(3)$ limit and is using a quadratic dependence
in $N_\nu$ for $\chi$ (\ref{chi2}).

Along this section we have learned to treat in a proper way 
binding energies (two-neutron separation energies) in transitional
regions where phase transitions might appear. We have seen that in
these regions one can get deviations from the linear tendency of
S$_{2n}$. Next step will be to find experimental examples where those
anomalies appear and explain them using the IBM. 
 
\section{Realistic calculations}
\label{real}
Within the framework of the IBM, the energy 
spectra and transition rates of
many medium-mass and heavy nuclei have been analyzed very
successfully. However, in most of these studies \cite{Cast88} 
the binding energies have been ignored. 
One of the reasons for such a neglect stems (partly) from the fact
that the early IBM studies (carried out in the 80's)
\cite{Scho80,Isac80,Bijk80} gave a rather good reproduction of the
general trends of the nuclear binding energy in different mass
regions. So, it was tacitly assumed that realistic IBM calculations
would also reproduce (in a natural way) 
the binding energy values. In the
mean time, with a large increase in the precision for the mass
measurement \cite{Kohl99,Schw98,Schw00}, much improved data have been
established \cite{Audi95}. Moreover, it must be noted that, in order
to carry out a proper comparison between calculated and experimental
binding energies (or $S_{2n}$ data), one cannot concentrate on single
or a few nuclei. Large chain of isotopes have to be considered and
this complicates considerably performing consistent IBM calculations.

In this section we will determine the experimental relations between
the control parameter $\xi$ ($\chi$) and the number of neutron pairs
$N_\nu$ through explicit fits of energy spectra for
selected nuclei in the three transitional regions. Starting from
these relations we are able to represent a specific chain 
of isotopes on the diagrams shown in section \ref{diagr}.
We subsequently show that the experimental binding energies (or 
two-neutron separation energies) are not always well reproduced 
although the spectra do. We analyze chains of isotopes in the
transitional regions (i) first using the more standard set of
parameters for the Hamiltonian \cite{Iach87}
and (ii), secondly, we compare with an 
alternative parametrization (see discussion below).  We do not aim a 
perfect fit of the experimental data, but we intend to emphasize the
influence of choosing various parametrizations on the binding
energies.  

\begin{figure}[]
\begin{center}
\mbox{\epsfig{file=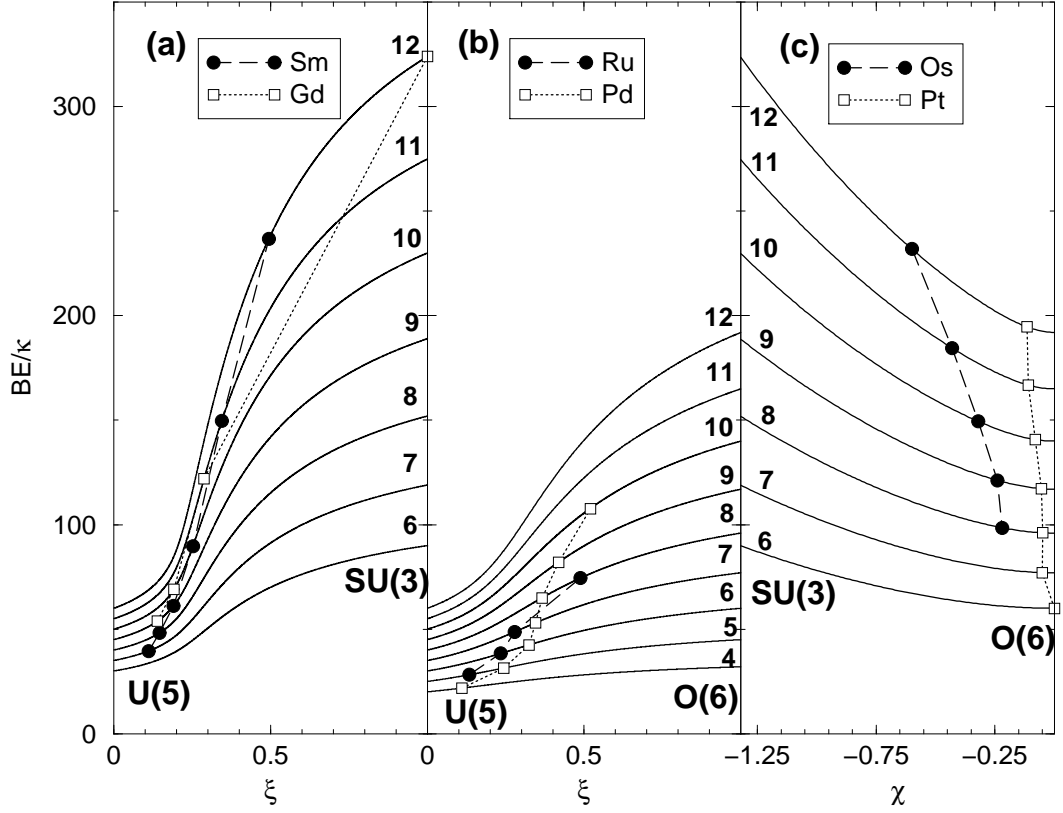,height=14.0cm,angle=-90}}
\end{center}
\caption{Binding energies in the laboratory system for different
numbers of bosons, $N$, as a function of the control parameter, 
in the transitional regions (a) $U(5)$-$SU(3)$, (b) $U(5)$-$O(6)$, 
and (c) $SU(3)$-$O(6)$. Full circles and open squares correspond to the
theoretical positions of different chains of isotopes.}
\label{fig-be-n}
\end{figure}

\begin{figure}[]
\begin{center}
\mbox{\epsfig{file=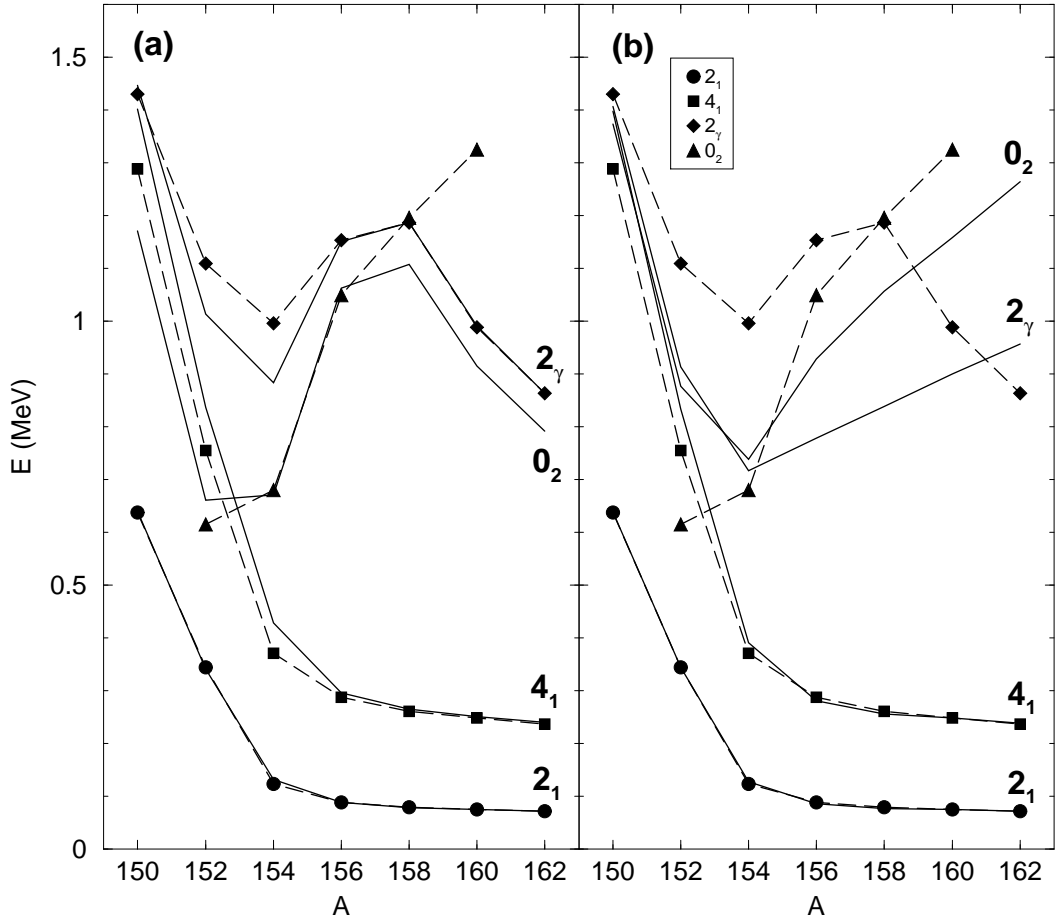,height=14.0cm,angle=-90}}
\end{center}
\caption{Comparison between theoretical and experimental energy levels
for Gd isotopes. Full lines correspond to IBM calculations, while
symbols connected by dashed lines correspond to experimental data. In
panels (a) and (b) different sets of parameters for the IBM
calculations are used, see text.}
\label{fig-Gd-ener}
\end{figure}

\begin{figure}[]
\begin{center}
\mbox{\epsfig{file=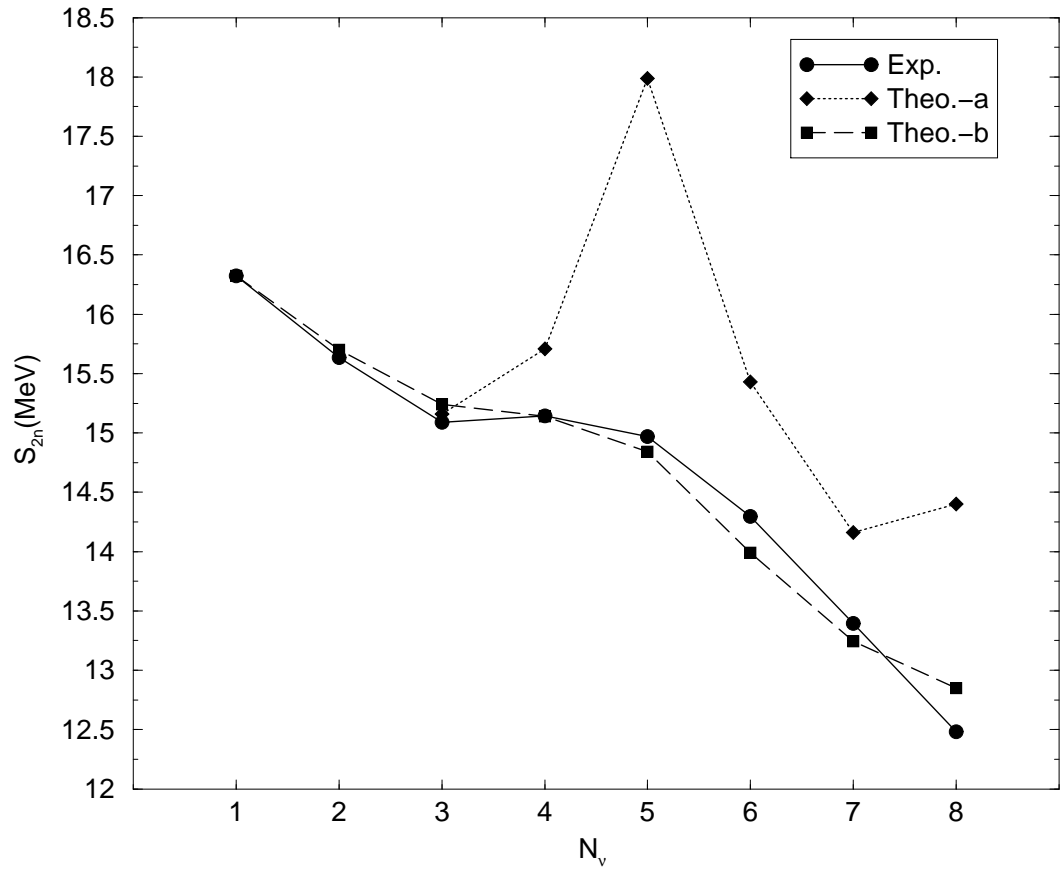,height=14.0cm,angle=-90}}
\end{center}
\caption{Two-neutron separation energies for Gd isotopes. 
Full lines correspond
to experimental data, while dotted and dashed lines correspond to two
theoretical calculations, see text.}
\label{fig-s2n-Gd}
\end{figure}

\begin{figure}[]
\begin{center}
\mbox{\epsfig{file=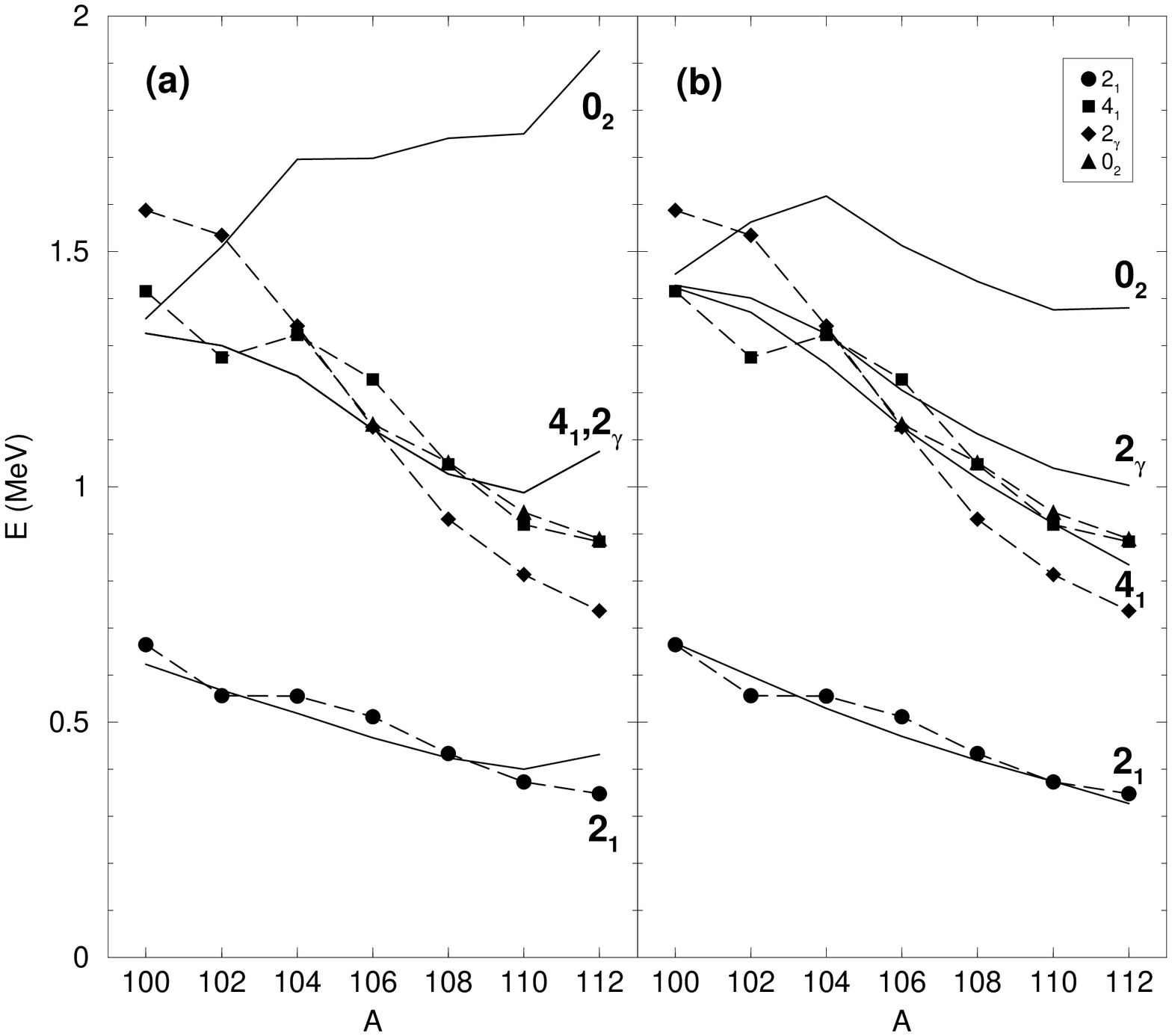,height=14.0cm,angle=-90}}
\end{center}
\caption{Same caption as in figure \ref{fig-Gd-ener} but for Pd isotopes.}
\label{fig-Pd-ener}
\end{figure}

\begin{figure}[]
\begin{center}
\mbox{\epsfig{file=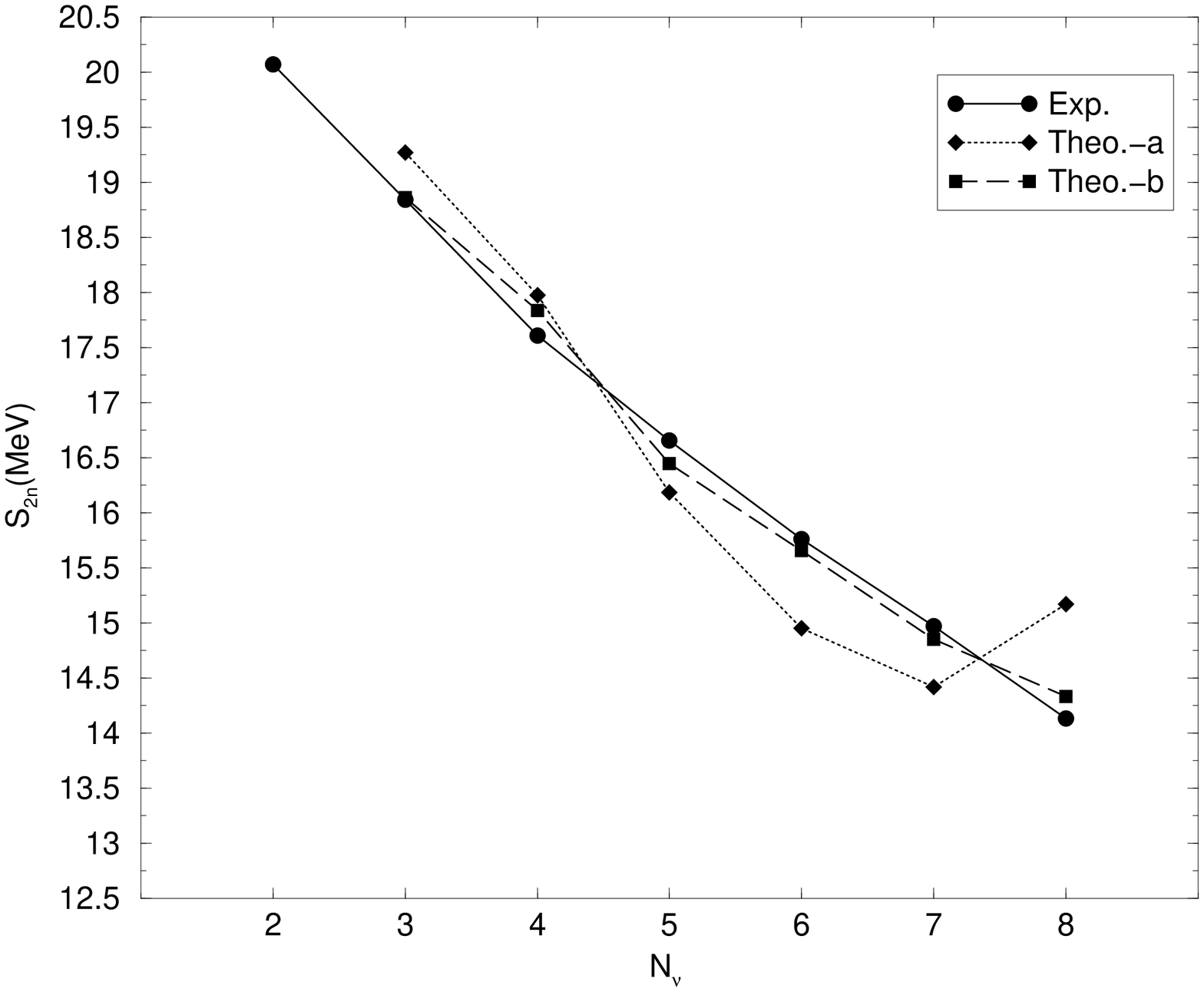,height=14.0cm,angle=-90}}
\end{center}
\caption{Same caption as in figure \ref{fig-s2n-Gd} but for Pd isotopes.}
\label{fig-s2n-Pd}
\end{figure}

Next we study the three transitional regions:

\begin{itemize}
\item \underline{The $U(5)$-$SU(3)$ transitional region}

There exit nuclei that exhibit a drastic change in the internal 
structure when one is passing their isotopes series, going 
from a vibrational into a rotational behavior. Typical examples are
situated in the rare-earth region, such as the Nd, Sm, and Gd nuclei. 
The Sm nuclei have been extensively studied in Ref. \cite{Scho78} and
a very good agreement with experimental data for S$_{2n}$ has been 
obtained. In figure \ref{fig-be-n}a, we plot the position of the 
different Sm isotopes and one can clearly see how
they cross the critical point, in particular $^{152}$Sm is situated
almost exactly where the phase transition happens. The cases of Nd and
Gd are very similar. Here, we will only focus to the case of Gd.

In this region we assume a value $\chi=-\sqrt{7}/2$ because some
Gd isotopes clearly exhibit the character of the $SU(3)$ dynamical 
symmetry \cite{Iach87}. This assumption was very 
successful in describing the Sm nuclei, 
which form neighboring nuclei. With this {\it ansatz} we have carried 
out a fit to the energy spectra of the Gd isotopes
(from $^{150}$Gd until $^{162}$Gd) using the Hamiltonian
(\ref{ham1}). In order to obtain the parameters, we tried to reproduce
the rotational structure of the ground- and $\gamma$-band as well as
possible. The $^{150}$Gd nucleus still shows a vibrational structure  
while $^{156-162}$Gd are considered as rather good $SU(3)$ 
examples. The parameters for $^{150}$Gd to $^{162}$Gd,
respectively, are shown in table \ref{tab-gd} (upper part).
In figure \ref{fig-Gd-ener}a we compare the theoretical and the 
experimental energies for the low-lying levels using this set of
parameters. Those
values are used to position the Gd isotopes in the phase
transition diagram of figure \ref{fig-be-n}a, where $^{150}$Gd
corresponds to $N=9$ while $^{156}$Gd corresponds 
to $N=12$. Note that
$^{158-162}$Gd are not plotted; they also corresponds to $\xi=1$. 
From figure \ref{fig-be-n}a one can observe a sudden transition 
in the Gd isotopes from a vibrational regime into the $SU(3)$ limit. 
Up to now, we have obtained a good description of the Gd energy
spectra but we have to consider the experimental two-neutron
separation energies too. This comparison is carried out in figure
\ref{fig-s2n-Gd} (the present calculation corresponds to {\it Theo.-a}),
using a linear contribution $S^{lin}_{2n}=15-0.252 N_\nu$ (MeV) as a
background. One observes a very bad agreement which points to the fact 
that the use of $\chi=-\sqrt{7}/2$ is not appropriate. A
different approach to fix the Hamiltonian is proposed in
\cite{Chou97}. There, it is assumed that in well deformed nuclei one
can keep the value of $\kappa$ constant around $\kappa=30$~keV and
smoothly change the values of $\chi$ and $\epsilon_d$ along a series of
isotopes. In the present case, the appropriate values for $\chi$
and $\kappa$ are $\chi=-0.6$ and $\kappa=19.2$~keV. 
On the other hand the values of
$\xi$ (using $N(1-\xi)/\xi=\epsilon_d/\kappa$, see equation
(\ref{ham1})) are shown in table \ref{tab-gd} (lower part)
in going on $^{146}$Gd until $^{162}$Gd (note
that in this case, a few extra isotopes are also considered). In figure
\ref{fig-Gd-ener}b we again compare the theoretical and the experimental
energies for the low-lying levels corresponding to the new
Hamiltonian. Again, we compare the calculated S$_{2n}$ values with 
the data and, as can be seen in figure
\ref{fig-s2n-Gd} (this calculation corresponds to {\it Theo.-b}),
using a linear contribution $S^{lin}_{2n}=16.75-0.607 N_\nu$ (MeV) as
a background, a very good agreement is obtained. Note that the
present results for $\xi=\xi(N)$ cannot be represented in 
figure \ref{fig-be-n}a due to the new value of $\chi$, since 
figure \ref{fig-be-n}a corresponds to $\chi=-\sqrt{7}/2$. It is worth
noting the strong similarities between figure \ref{fig-s2n-Gd} and
figure \ref{fig-s2n-u5-su3}b.

\item \underline{The $U(5)$-$O(6)$ transitional region}

Clear-cut examples of nuclei in which the structure changes from 
spherical to $\gamma$-unstable are obtained in the Ru and Pd region. 
Both series of isotopes have already been studied in \cite{Stac82}, 
showing that the lighter isotopes exhibit a
vibrational structure while the heavier ones present a
$\gamma$-unstable 
behavior. In reference \cite{Stac82} the general trends have been 
analyzed but it was
shown that difficulties appear when trying to obtain a consistent
description for several observables at the same time. 
Due to the strong similarities between both nuclei, only the study of 
Pd nuclei will be presented here.

A first assumption will be to consider $\chi=0$, which is suggested
in reference \cite{Stac82} too. Using the Hamiltonian (\ref{ham1}) we
fit the energy spectra of $^{100-112}$Pd. No other isotopes 
are fitted because
only for those the energy ratio $E(4_1^+)/E(2_1^+)$ changes between $2$
($U(5)$) and $2.5$ ($O(6)$). Note that this energy ratio will be the main
observable to be reproduced during the fits. We have taken 
a fixed single particle energy at $\epsilon_d=65$~keV 
and the parameters 
that we have obtained are shown in table \ref{tab-pd} (upper part).
Note that no rotational term is necessary. 
In figure \ref{fig-Pd-ener}a we compare the theoretical and the 
experimental low-lying levels for this set of parameters.
Those values are used
to position the Pd isotopes on figure \ref{fig-be-n}b, where
$^{100}$Pd corresponds to $N=4$ and $^{112}$Pd 
corresponds to $N=10$. One can
clearly see that the Pd isotopes cover only half the range
between the $U(5)$ and the $O(6)$ limit. In figure \ref{fig-s2n-Pd}
one obtains a poor description when comparing S$_{2n}$ (this
calculation corresponds to {\it Theo.-a}), with a linear contribution 
$S^{lin}_{2n}=22.27-1.295 N_\nu$ (MeV) as background, while the energy
spectra are well reproduced. 
Again, this suggests a change in the value of $\chi$. In the
following, we will use $\chi=-0.3$, a value that produces a realistic
description of the energy spectra. 
In this case one can fix the single particle energy
to $\epsilon_d=70$~keV. The obtained values of $(\kappa,\xi)$ from 
$^{100}$Pd until $^{114}$Pd  are shown in table \ref{tab-pd} 
(lower part). 
In figure \ref{fig-Pd-ener}b we compare again
the theoretical and the experimental low-lying levels for the
new Hamiltonian. Now, however, one obtains
good agreement when comparing
experimental and theoretical S$_{2n}$ values (this calculation
corresponds to {\it Theo.-b}), with a linear contribution 
$S^{lin}_{2n}=20.46-0.835 N_\nu$ (MeV) as a background,
as can be seen in figure
\ref{fig-s2n-Pd}. The new results cannot be represented on figure
\ref{fig-be-n}b because $\chi\neq0$.

\item \underline{The $SU(3)$-$O(6)$ transitional region}

The most clear-cut examples of nuclei situated in this transitional 
region are the Os and Pt nuclei. The lighter isotopes exhibit rotational
structures while the heavier ones come very close to the
$\gamma$-unstable limit. This region has already been studied in
\cite{Cast78} in the framework of the IBM-1 but using a
different parametrization for the Hamiltonian. 
Due to the similarities among both series of isotopes, only the 
analysis of the Pt nuclei will be presented here.

In order to make a global fit we take into account the fact that
$^{196}$Pt is one of the best examples of a $\gamma$-unstable nucleus
and, as a consequence, $\chi=0$ and $\xi=0$ \cite{Iach87}. 
The obtained parameters
$(\kappa,\chi,\kappa')$ for $^{184}$Pt until $^{196}$Pt are shown 
in table \ref{tab-pt} (upper part).
In figure \ref{fig-Pt-ener}a we compare the theoretical and the 
experimental low-lying levels for these parameters. 
If one places these values in figure \ref{fig-be-n}c, where $^{196}$Pt 
corresponds to $N=6$ while $^{184}$Pt corresponds to $N=12$, 
it is clear that the Pt isotopes always stay very
close to the $O(6)$ limit. Finally, in figure \ref{fig-s2n-Pt}, we 
compare the theoretical and experimental S$_{2n}$ values (this
calculation corresponds to {\it Theo.-a}), with a linear contribution
$S^{lin}_{2n}=12.66-0.545 N_\nu$ (MeV) as a background. Again, we
obtain a poor agreement in the description of the experimental
values. In order to treat this mass region with a slightly different  
Hamiltonian, one can follow again Ref.~\cite{Chou97}. There, it is 
suggested to  
include a single-particle energy term. In the case of the Pt
nuclei the more appropriate value is $\epsilon_d=25.7$~keV. The value 
of $\kappa$ is also fixed to $\kappa=33.5$~keV and
$\kappa'=15.2$~keV. The different values
of the parameters in the Hamiltonian for $^{184}$Pt until 
$^{196}$Pt are shown 
in table \ref{tab-pt} (lower part).
In figure \ref{fig-Pt-ener}b we know compare the theoretical and the 
experimental low-lying levels for the new Hamiltonian. 
Using this new parametrization, when comparing the
S$_{2n}$ values (this calculation corresponds to {\it Theo.-b}), with a
linear contribution $S^{lin}_{2n}=11.85-0.649 N_\nu$ (MeV) as a
background, a good agreement is obtained as can be seen in figure
\ref{fig-s2n-Pt}. These new results cannot be represented in figure
\ref{fig-be-n}c because now $\epsilon_d\neq0$.
\end{itemize}

In this section we have presented two different parameter sets
obtained from fitting  
the energies of the low-lying states of different isotopes of Gd, Pd, and
Pt. A very striking fact is, on one side, the good description of
the energy spectra for both sets of parameters, and on the other side, 
the very different results obtained for the two-neutron
separation energy using these different parametrizations. 
The results obtained here were totally unexpected
because it was assumed tacitly that a good description of energy
spectra would imply a good description for the binding energies too. 
In order to illustrate this problem, we
consider the case of $^{148}$Sm which has been studied by different
authors \cite{Scho78,Chou97}. Although in all the cases the
description of the energy spectrum is
satisfactory, when calculating the binding energy for the
proposed Hamiltonians, one obtains the results $0.939$~MeV
(Ref~\cite{Scho78}),
$1.571$~MeV (Ref~\cite{Chou97}) and $2.253$~MeV (also in
Ref~\cite{Chou97}). All this suggests that a consistent study of 
atomic nuclei, within the IBM, requires the inclusion of an analysis 
of the binding energy. 

\section{Summary and conclusions}
\label{conclu}
In the present article,  we have analyzed  transitional regions and phase
transitions in the framework of the IBM. A new kind of diagram (see
figure \ref{fig-be-n}) for dealing with phase transitions has been 
presented. In this diagram we plot the binding energy of the system 
versus the control parameter (Hamiltonian) for a range in the number of
bosons. This procedure turns out to be a particularly convenient
method when working in finite systems without a true control parameter.  

The main conclusions are twofold:

\noindent
(i) We have shown that, although the crossing through the phase
transition points produces a deviation in the S$_{2n}$ values from the
overall linear background dependence, the way in which, in the given
chain of isotopes, this critical point is crossed proves extremely
important in deriving clear-cut results. If {\it e.g.~}the relation
between a control parameter and the isotope location 
({\it e.g.~}boson number) is linear, one does not 
observe dramatic changes in
S$_{2n}$ in any of the three transitional regions ($U(5)$-$SU(3)$,
$U(5)$-$O(6)$, and $SU(3)$-$O(6)$). On the contrary, for a quadratic
relation, clear changes can appear in the S$_{2n}$ behavior at the
crossing point.   

\noindent
(ii) We have observed that in a consistent study of long chains of
isotopes, one has to treat the ground-state 
(through its binding energy) 
on equal level with the excited states (relative energy spectrum). 
It turns out that parameters producing very similar energy spectra 
can still result in important differences in the ground-state 
binding energy (order of $1$ MeV).

\section{Acknowledgments}
The authors are grateful to G.~Bollen and co-workers, 
R.F.~Casten, F.~Iachello, J.~Jolie, P.~Van Isacker, and 
J.~Wood for stimulating discussions.
The authors like to thank the ``FWO-Vlaanderen'' for financial
support.

\begin{figure}[]
\begin{center}
\mbox{\epsfig{file=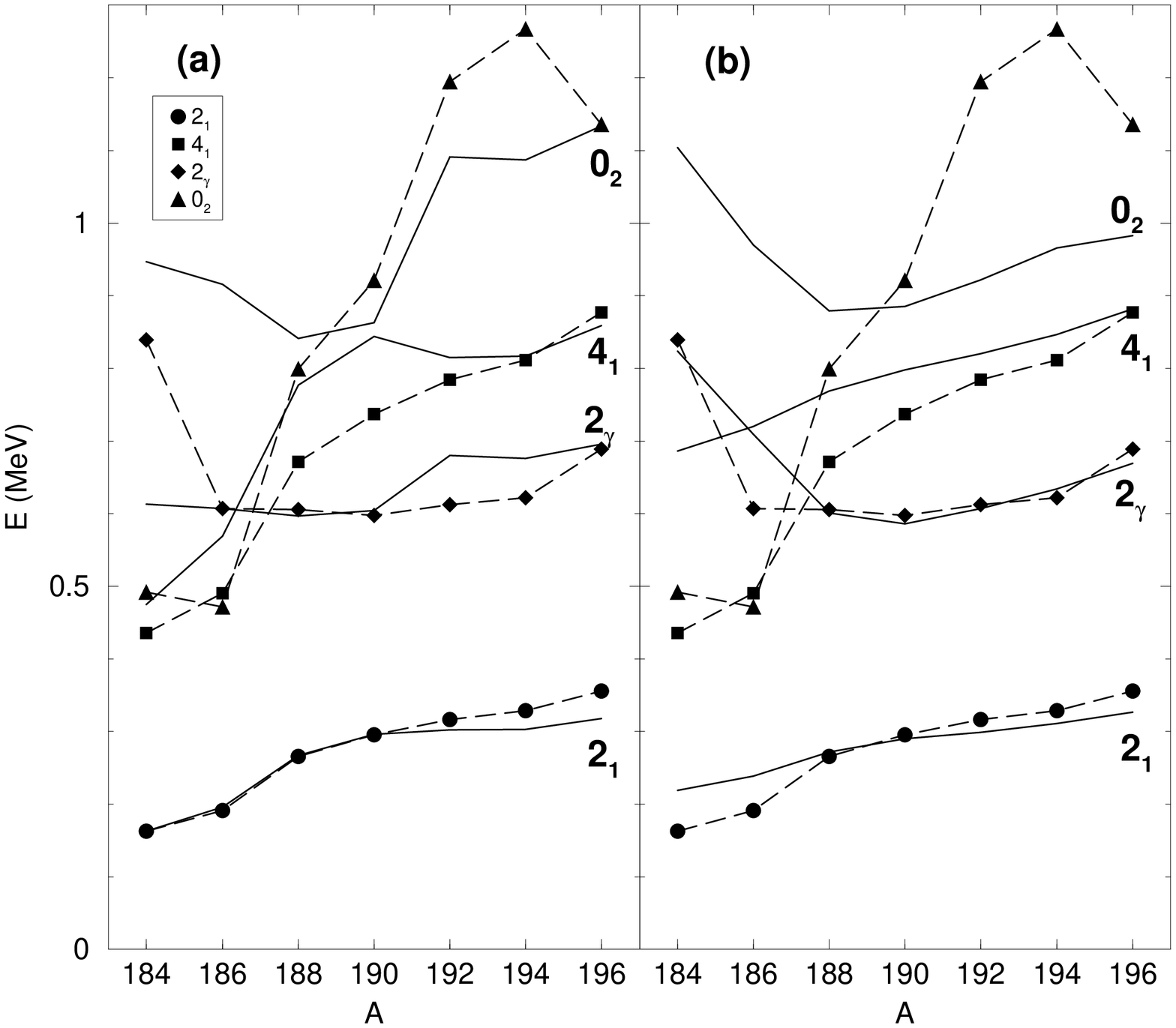,height=14.0cm,angle=-90}}
\end{center}
\caption{Same caption as in figure \ref{fig-Gd-ener} but for Pt isotopes.}
\label{fig-Pt-ener}
\end{figure}

\begin{figure}[]
\begin{center}
\mbox{\epsfig{file=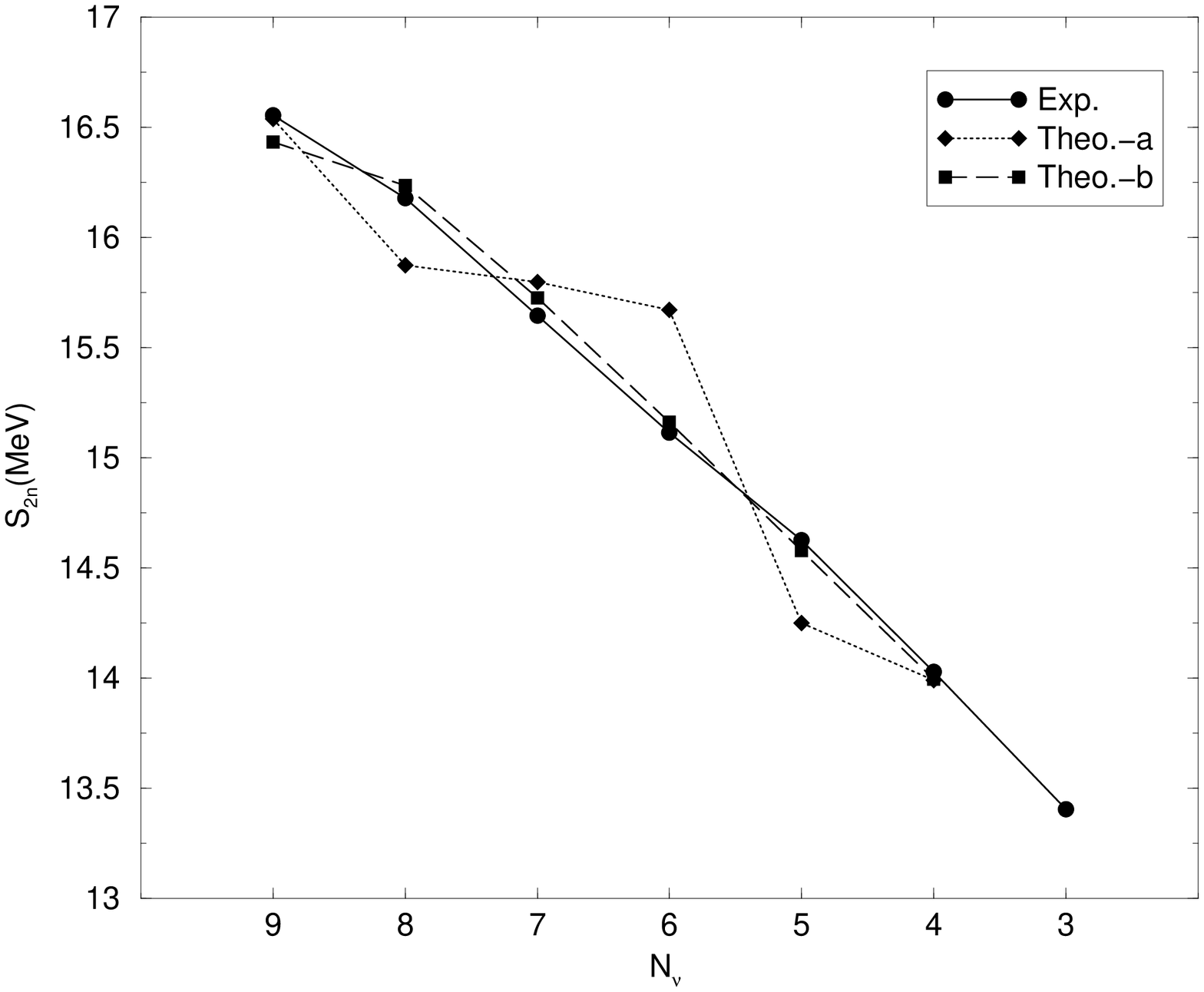,height=14.0cm,angle=-90}}
\end{center}
\caption{Same caption as in figure \ref{fig-s2n-Gd} but for Pt isotopes.}
\label{fig-s2n-Pt}
\end{figure}

\begin{table}
\caption{Parameters of the Hamiltonian for Gd isotopes ($N_\pi=7$). 
Upper part of
the table is referred as {\it Theo.-a} and lower part as {\it Theo.-b}
(as used in the text and in the figures).} 
{\it Theo.-a}
\begin{tabular}{cccccccc}
$A$      &150   &152   &154   &156   &158   &160   &162  \\
$N_\nu$  &2     &3     &4     &5     &6     &7     &8    \\
\hline
$\kappa$ &15.4  &15.4  &15.4  &15.4  &14.8  &11.3  &9.1 \\
$\xi$    &0.139 &0.192 &0.287 &1     &1     &1     &1   \\
$\kappa'$&9.0   &9.0   &9.0   &9.0   &7.7   &8.3   &8.6 \\
\end{tabular}
$\kappa$, $\kappa'$ in keV and $\xi$ dimensionless, $\chi=-\sqrt{7}/2$.

{\it Theo-b}
\begin{tabular}{cccccccccc}
$A    $  &146   &148   &150   &152   &154   &156   &158   &160   &162 \\
$N_\nu$  &0     &1     &2     &3     &4     &5     &6     &7     &8   \\
\hline
$\xi$    &0.60  &0.137 &0.166 &0.236 &0.373 &0.535 &0.625 &0.658 &0.724
\end{tabular}
$\kappa=19.2$ keV, $\kappa'=0$ and $\xi$ dimensionless, $\chi=-0.6$.
\label{tab-gd}
\end{table}

\begin{table}
\caption{Parameters of the Hamiltonian for Pd isotopes ($N_\pi=2$). 
Upper part of
the table is referred as {\it Theo.-a} and lower part as {\it Theo.-b}
(as used in the text and in the figures).} 
{\it Theo.-a}
\begin{tabular}{cccccccc}
$A$      &100   &102   &104   &106   &108   &110   &112  \\
$N_\nu$  &2     &3     &4     &5     &6     &7     &8    \\
\hline
$\kappa$ &20.0  &42.0  &52.0  &49.0  &47.0  &52.0  &70.0 \\
$\xi$    &0.110 &0.244 &0.324 &0.345 &0.366 &0.419 &0.52 \\
\end{tabular}
$\kappa$ in keV, $\kappa'=0$ and $\xi$ dimensionless, $\chi=0$.

{\it Theo-b}
\begin{tabular}{cccccccccc}
$A$      &100   &102   &104   &106   &108   &110   &112   &114 \\
$N_\nu$  &2     &3     &4     &5     &6     &7     &8     &9   \\
\hline
$\kappa$ &22.0  &44.0  &50.0  &44.0  &40.0  &37.0  &37.0  &33.0 \\
$\xi$    &0.112 &0.239 &0.300 &0.306 &0.314 &0.322 &0.346 &0.342
\end{tabular}
$\kappa$ in keV, $\kappa'=0$  and $\xi$ dimensionless, $\chi=-0.3$.
\label{tab-pd}
\end{table}

\begin{table}
\caption{Parameters of the Hamiltonian for Pt isotopes ($N_\pi=2$). 
Upper part of
the table is referred as {\it Theo.-a} and lower part as {\it Theo.-b}
(as used in the text and in the figures).} 
{\it Theo.-a}
\begin{tabular}{cccccccc}
$A$      &184    &186    &188    &190    &192    &194    &196   \\
$N_\nu$  &10     &9      &8      &7      &6      &5      &4     \\
\hline
$\kappa$ &43.0  &44.0  &44.0  &47.0  &60.0  &60.0  &63.0 \\
$\chi$   &-0.115&-0.110&-0.080&-0.055&-0.049&-0.050&0    \\
$\kappa'$&4.2   &0.8   &17.6  &19.0  &11.0  &11.0  &11.0 
\end{tabular}
$\kappa$ and $\kappa'$ in keV and $\xi$ dimensionless,
$\varepsilon_d=0$.  

{\it Theo-b}
\begin{tabular}{cccccccccc}
$A$      &184    &186    &188    &190    &192    &194    &196   \\
$N_\nu$  &10     &9      &8      &7      &6      &5      &4     \\
\hline
$\kappa$ &33.5   &33.5   &33.5   &33.5   &33.5   &33.5   &33.5 \\
$\chi$   &-0.25  &-0.20  &-0.10  &0      &0      &0      &0    \\
$\kappa'$&15.2   &15.2   &15.2   &15.2   &15.2   &15.2   &15.2 
\end{tabular}
$\kappa$ and $\kappa'$ in keV and $\xi$ dimensionless,
$\varepsilon_d=25.7$ keV.   

\label{tab-pt}
\end{table}
\end{document}